\documentclass[a4paper]{aa}

\usepackage{xspace}
\usepackage{graphicx}
\usepackage{txfonts}
\usepackage{natbib}
\usepackage{lscape}
\usepackage[dvipdfm,breaklinks=true,bookmarks=true,hypertexnames=false,bookmarksnumbered=true,bookmarksopen=false,bookmarks=true,bookmarksnumbered=true]{hyperref}

\newcommand{\msun}{{\rm M}_{\odot}}
\newcommand{\lsun}{{\rm L}_{\odot}}
\newcommand{\kms}{\, {\rm km\, s}^{-1}}
\newcommand{\h}{\,h_{70}}
\newcommand{\hmm}{\h^{-1}}

\newcommand{\kpc}{\, {\rm kpc}}

\newcommand{\hmkpc}{\hmm\kpc}
\newcommand{\hmlsun}{\h^{-2}\, \lsun}

\newcommand{\Mpc}{\, {\rm Mpc}}

\newcommand{\der}{{\rm d}}
\newcommand{\kev}{{\rm keV}}

\newcommand{\eg}{{\it e.g.}\xspace}

\newcommand{\zl}{z_{\rm lens}}
\newcommand{\zs}{z_{\rm arc}}
\newcommand{\zrad}{z_{\rm dbl}}
\newcommand{\Reff}{R_{\rm e}}
\newcommand{\data}{{\rm data}}
\newcommand{\lname}{SL2SJ02176-0513}
\newcommand{\mypm}[2]{^{+#1}_{-#2}}

\newcommand{\ringfinder}{{\tt ringfinder}}
\newcommand{\lenstool}{{\tt lenstool}}
\newcommand{\hyperz}{{\tt hyperz}}

\begin{document}

\title{
  The mass profile of early-type galaxies in overdense environments: the case of the double source plane gravitational lens \lname \thanks{
    Based on observations obtained with MegaPrime/MegaCam, a
    joint project of CFHT and CEA/DAPNIA, at the Canada-France-Hawaii
    Telescope (CFHT) which is operated by the National Research
    Council (NRC) of Canada, the Institut National des Science de
    l'Univers of the Centre National de la Recherche Scientifique
    (CNRS) of France, and the University of Hawaii. This work is based
    in part on data products produced at TERAPIX and the Canadian
    Astronomy Data Centre as part of the Canada-France-Hawaii
    Telescope Legacy Survey, a collaborative project of NRC and
    CNRS.}
}

\author{
  H.~Tu\inst{1,2,3},
  R.~Gavazzi\inst{2,3,4},
  M.~Limousin\inst{5,6,7},
  R.~Cabanac\inst{5},
  P.~J.~Marshall\inst{4},
  B.~Fort\inst{2,3},
  T.~Treu\inst{4},
  R.~P\'{e}llo\inst{8},
  E.~Jullo\inst{6,9},
  J.-P.~Kneib\inst{6,10},
  J.-F.~Sygnet\inst{2,3}
}

\institute{
  Physics Department \& Shanghai Key Lab for Astrophysics, Shanghai
  Normal University, 100 Guilin Road, Shanghai 200234,  China \\
  \email{tuhong@shnu.edu.cn}
  \and
  CNRS, UMR7095, Institut d'Astrophysique de Paris, 98bis Bd Arago, F-75014, Paris, France
  \and
  UPMC Univ. Paris 6, UMR7095, Institut d'Astrophysique de Paris, 98bis Bd Arago, F-75014, Paris, France
  \and
  Department of Physics, University of California, Santa Barbara, CA 93106, USA
  \and
  Laboratoire d'Astrophysique de Toulouse-Tarbes, Universit\'e de Toulouse, CNRS,
  57 avenue d'Azereix, F-65000 Tarbes, France
  \and
  OAMP, Laboratoire d'Astrophysique de Marseille - UMR 6110 - Traverse du siphon, F-13012 Marseille, France
  \and
  Dark Cosmology Centre, Niels Bohr Institute, University of Copenhagen,
  2100 Copenhagen, Denmark
  \and
  Laboratoire d'Astrophysique de Toulouse-Tarbes, CNRS-UMR 5572, and Univ. Paul Sabatier, F-31400 Toulouse, France
  \and
  European Southern Observatory, Alonso de Cordoba, Santiago, Chile
  \and
  Department of Astronomy, California Institute of Technology, 105-24, Pasadena, CA91125, USA
}



\abstract
{ The Strong Lensing Legacy Survey (SL2S) provides a sample of strong
lensing events associated with massive distant galaxies, some of which
lie in the outskirts of galaxy groups and clusters.  }
{ 
We investigate the internal structure of early-type galaxies in
overdense environments, where tidal forces are expected to alter dark
matter halos of infalling galaxies.
}
{\lname~is a remarkable lens
for the presence of two multiply-imaged systems at different redshifts
lensed by a foreground massive galaxy at $\zl=0.656$: a bright cusp
arc at $\zs=1.847$ and an additional double-image system at an
estimated redshift of $\zrad\sim2.9$ based on photometry and lensing
geometry. The system is located about 400~kpc away from the center of a
massive group of galaxies. Mass estimates for the group are available
from X-ray observations and satellite kinematics. Multicolor
photometry provides an estimate of the stellar mass of the main lens
galaxy. The lensing galaxy is modeled with two components (stars and
dark matter), and we include the perturbing effect of the group
environment, and all available constraints.}
{ We find that classic lensing degeneracies, e.g. between external
  convergence and mass density slope, are significantly reduced with
  respect to standard systems and infer tight constraints on the mass
  density profile: (i) the dark matter content of the main lens galaxy
  is in line with that of typical galaxies $f_{\rm
  dm}(<\Reff)=0.41\mypm{0.09}{0.06}$; (ii) the required mass
  associated with the dark matter halo of the nearby group is
  consistent with X-ray and weak-lensing estimates
  ($\sigma_{\rm grp}=550\mypm{130}{240}$); 
  (iii) accounting for the group contribution in the form of an external
  convergence, the slope of the mass density profile of the main lens
  galaxy alone is found to be $\alpha=-1.03\mypm{0.22}{0.16}$,
  consistent with the isothermal ($\alpha=-1$) slope.}
{ We demonstrate that multiple source plane systems together with good
  ancillary dataset can be used to disentangle local and environmental
  effects.}


\authorrunning{Tu et al.}
\titlerunning{The double lens plane system \lname }

\maketitle


\section{Introduction}

In the last thirty years, gravitational lensing has become a major
astrophysical tool because of its unique ability to probe the densest
regions of the universe independent of their emitted radiation. Large
imaging or spectroscopic surveys are becoming key players, in that
context, by bringing numerous lensing events of all scales and
spanning an increasingly broader range of masses and redshifts
\citep[\eg][]{kochanek99,myers03b,inada08,bolton06,bolton08a,faure08a,marshall08}.

Among the imaging searches for lenses, the CFHT Legacy
Survey\footnote{\url{http://www.cfht.hawaii.edu/Science/CFHLS/}}
stands out by combining depth in five photometric bands, excellent
image quality ($\sim0\farcs7-0\farcs8$ median seeing) and an area of
170 deg$^2$ when completed. The Strong Lensing Legacy Survey
(SL2S)\footnote{\url{http://www-sl2s.iap.fr/}} aims at automatically
detecting gravitational lensing events in these data
\citep{cabanac07,limousin08}, be they either large arcs in groups and
clusters, or smaller, galaxy-scale lensing events (sometimes also
referred to as ``Einstein rings'' to remind us that the lensed
background source is extended, and therefore does, at some surface
brightness level, produce a complete ring around the deflector). A
detailed description of the \ringfinder~detection method will be given
in a forthcoming paper (Gavazzi et al.~2009, in prep).

One of the scientific questions that can be addressed in a direct and
powerful manner by gravitational lensing is that of the dark matter
halo profiles of early-type galaxies, the most common type of strong
lenses. For example, gravitational lensing has been used to
demonstrate the presence of dark matter halos
\citep[\eg][]{kochanek94,treu04}, and to test the validity of the
universal dark matter halos predicted by numerical simulations
\citep[][]{navarro1997,gavazzi07}. When galaxies fall into an
overdense environment, i.e. a cluster or a group, their dark matter
halos are expected to be stripped by tidal interactions
\citep[\eg][]{dobke07}. Tidal stripping typically does not reach deep
enough into the infalling galaxy to affect its luminous structure
\citep[e.g][]{treu03} until the galaxy reaches the densest regions,
and therefore the effects are hardly detectable with classical
astronomical probes. However, stripping of the outer halo begins much
earlier during the infall and can be detected by gravitational
lensing, for example by weak lensing analysis of cluster galaxies
\citep[][]{natarajan97,geiger98,natarajan02,natarajan07}. Analyses of selected
galaxy samples coupling lensing and dynamics have also suggested
that halos of satellite galaxies may be truncated and effectively
steeper than those of isolated galaxies
\citep[\eg][]{treu02,auger08,treu08}. In turn, galaxy-scale strong
lenses in the field of clusters, can be used to constrain the cluster
mass distribution near the high magnification regions of the cluster
\citep[][]{tu08,limousin08} and that of nearby structures
\citep[\eg][]{kochanek91,king07,tu08}.

Beyond the general problem of tidal truncation of dark matter halos,
the environment of lens galaxies has been a topic of intense debate in
the lensing literature for a variety of reasons. One important
question is the role of the environment in the measurement of Hubble's
constant from lens time delays \citep[\eg][]{kochanek02,kochanek04b}.
In addition, the environment of lens galaxies must be understood if
one wants to generalize the findings to the overall population of
similar galaxies. It is known that multiply-imaged QSOs are often
found behind overdense regions which have noticeable effects on the
lensing configurations
\citep[\eg][]{keeton97,fassnacht02,auger07,momcheva06,oguri06,williams08}.
This preference for dense environments is also seen in the
low-redshift SLACS sample \citep{treu08}. However, these authors
showed that this preference is no more pronounced than for any massive
elliptical galaxy, whether or not it is actually a lens. Within the
errors, the preference for high-density environments is just a
consequence of the clustering of massive galaxies. A somewhat similar
trend is observed in the COSMOS survey \citep{faure08b} although the
use of weak lensing convergence map as a proxy for the environment
makes the interpretation a bit more difficult. This preference for
clustering is nicely illustrated by the results of \citet{fassnacht06}
and \citet{newton08} which can be summarised as: {\it a good place to
look for new lenses is the vicinity of known lenses}.

In this paper, we present the analysis of an extraordinary system
\lname\, discovered by the CFHT Legacy Survey, and
use it as a case study to investigate the effect of environment on the
dark matter halos of galaxies falling into overdense environment. The
system is unusual for several reasons. First, it is producing multiple
images of two sources at different redshifts, similarly to the double
Einstein Ring discovered by \citet[][]{gavazzi08}.  Second, the system
is located near a well-constrained group environment, and we were able
to gather a large spectrophotometric set of information for both the
lensing galaxy and its environment. Using this wealth of information,
we are able to break some of the classic degeneracies of lensing
studies. This allows us to infer the relative distribution of light
and dark matter in the lens galaxy out to 1.5~effective radii of the
lens galaxy, and probe the effect of the nearby group potential.

The paper is organised as follows. After presenting the data set
available for \lname~in Sect.\ \ref{sec:data}, we measure the optical
properties of the lens galaxy (including stellar mass) in Sect.\
\ref{sec:LensPpty}, and then present our lens modelling in Sect.\
\ref{sec:model}. In Sect.\ \ref{sec:modelComp}, we discuss the main
conclusions from the modelling and future possible improvements.

Unless otherwise stated we assume a concordance cosmology with
$H_0 = 70\h\,\kms\Mpc^{-1}$, $\Omega_{\rm m}=0.3$ and $\Omega_\Lambda=0.7$.
At the redshift of the lens \lname~$\zl=0.6459$, $1\arcsec\sim 6.9\kpc$.
All magnitudes are expressed in the AB system.


\section{Observational data}\label{sec:data}

In this section we present the ensemble of available data.
Table~\ref{tab:data} summarises the main quantities of interest for this work.

\begin{figure*}[htb]
  \centering
  \includegraphics[width=\hsize]{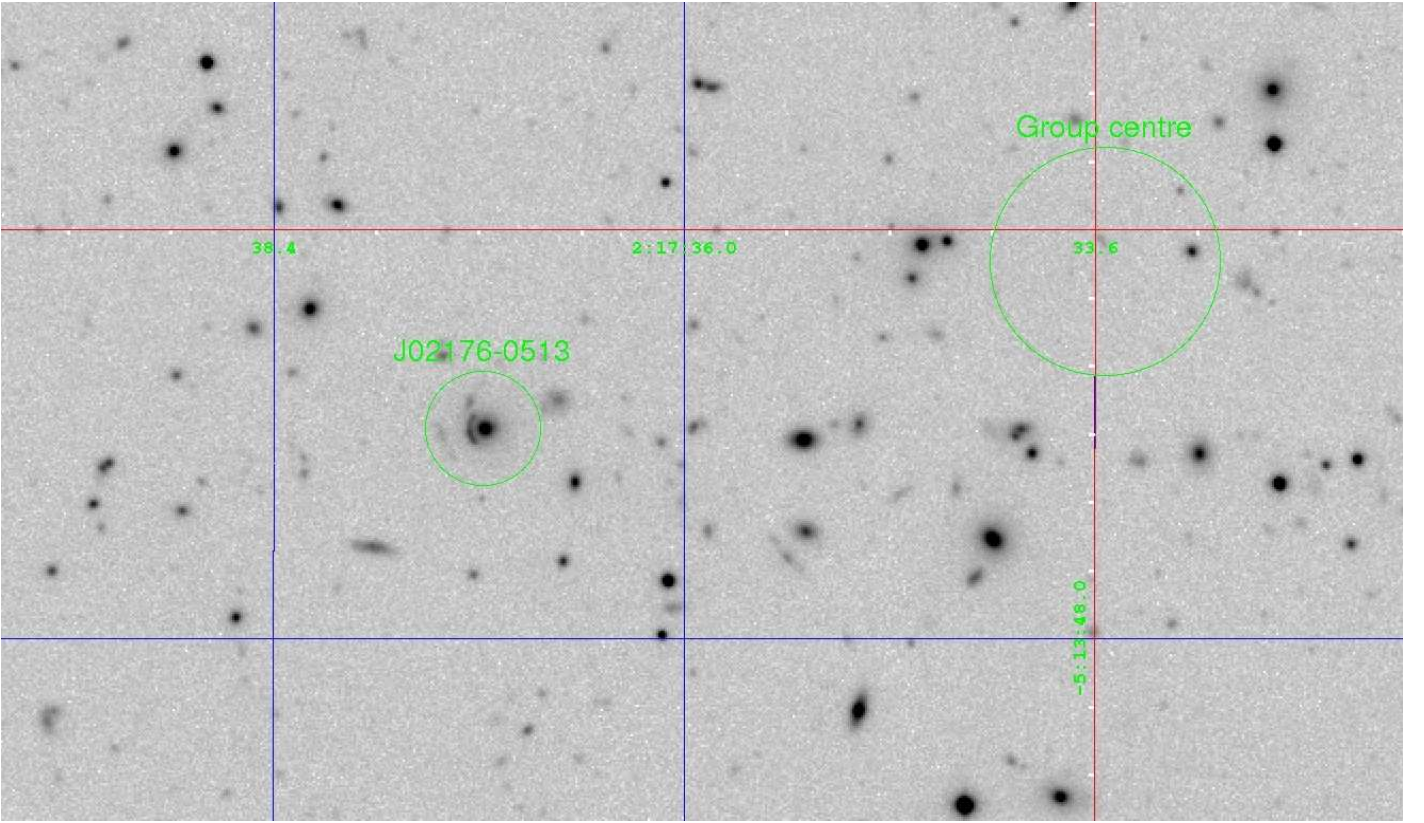}
  \caption{Wide field CFHT image of \lname~($irg$ ``inverted'' composite image). The big 10$\arcsec$ radius circle shows the location of the nearby galaxy group studied by \citet{geach021737}.}\label{021737cfht}
\end{figure*}

\begin{table}
  \caption{Summary of the most relevant observational quantities for \lname.
The ellipticity is defined as  $\epsilon=(a^2-b^2)/(a^2+b^2)$. Masses are given with their photometric errors. References: $^a$ \citet{geach021737}.
Note that errors on photometry are dominated by systematic calibration uncertainties.}
  \label{tab:data}
  \centering
  \begin{tabular}[h!]{lcc}
    \hline\hline
    Lens galaxy centre (J2000) & RA  & 02:17:37.136\\
                               & DEC  &-05:13:29.41 \\

    Lens galaxy redshift        & $\zl$ &  $0.6459 \pm 0.0003$ \\
    Lens mags ($R<1\arcsec$)     &  u*  & $23.90 \pm 0.20$\\
                                 &  g'  & $22.89 \pm 0.07$\\
                                 &  r'  & $21.74 \pm 0.05$\\
                                 &  i'  & $20.44 \pm 0.04$\\
                                 &  z'  & $20.12 \pm 0.03$\\
     HST (S\'ersic model mags)   &  F606W  & $ 20.90 \pm 0.05$\\
                                 &  F814W  & $ 19.71 \pm 0.01$\\
     Effective radius (F814W)    &  $\Reff$  & $ 0\farcs90 \pm 0\farcs10 $ \\
                                 &         & $ 6.20 \pm 0.7 \hmkpc $ \\
     S\'ersic model index (F814W)&   n     & $ 5.20$ \\
     Rest frame V band luminosity& $L_V$   & $1.36\mypm{0.08}{0.08}  \times 10^{11} \hmlsun$\\
     light position angle (F814W)&  $PA_*$  & $ 78\pm 6$ deg \\
     light ellipticity (F814W)   &  $\epsilon_*$  & $ 0.14\pm 0.06 $ \\
     {\it ``Photometric''} stellar mass &  $M_*$ & $ 2.1-4.1 \times 10^{11}\msun$ \\

\hline
     Tangential arc &&\\
        $\ldots$  redshift  & $\zs$ & $1.8470 \pm 0.0003$\\
        $\ldots$            & F606W$_{\rm{arc}}$ & $22.04 \pm 0.16 $ \\
        $\ldots$            & F814W$_{\rm{arc}}$ & $22.19 \pm 0.13 $ \\
\hline
    Double system (R1) &&\\
        $\ldots$    photom. redshift& $\zrad$ &$2.90\mypm{0.18}{0.24}$ \\
        $\ldots$                    & F606W$_{\rm R1}$ & $24.11 \pm 0.06 $ \\
        $\ldots$                    & F814W$_{\rm R1}$ & $23.91 \pm 0.06 $ \\
\hline
    Einstein radius of the main arc& $R_{arc}\,\,(\arcsec)$ & 1.4 \\
\hline
    Group redshift$^a$  &  $z_{\rm grp} $  & $0.648\pm0.001$ \\
    Group velocity dispersion$^a$  &  $\sigma_{\rm grp}^{\rm spec} $  & $770 \pm 170 \kms$ \\
    X-ray temperature$^a$ &  $ kT $  & $2.0\mypm{1.0}{0.6}$ keV \\
    $\sigma$ from $\sigma-T_X$ relation$^a$  &  $\sigma_{\rm grp}^{\rm X}$  & $520 \pm 120 \kms$ \\
\hline
  \end{tabular}
\end{table}


\subsection{\lname~in the CFHTLS imaging data}\label{sec:data:survey}

The SL2S detection strategy of lensing events was introduced by
\citet{cabanac07}. Details of the \ringfinder~detection algorithm as well
as the full sample of gravitational Einstein rings will be presented in a
forthcoming paper (Gavazzi et al.\ 2009, in prep.). As part of this effort the
gravitational lens \lname~was detected as a promising galaxy-scale lens
candidate. Subsequent high-resolution follow up imaging and spectroscopy were
since obtained. \lname~belongs to the W1 field in the {\it wide} part of the
CFHTLS survey. Currently, sub-arcsecond seeing imaging is available in the
Megacam bands u*, g', r', i' and z'. A large scale colour image of \lname~is
shown in Fig.~\ref{021737cfht}. The coordinates of the elliptical lens on the
CFHTLS image is $\alpha_{J2000}$=02:17:37.136, $\delta_{J2000}$=-05:13:29.41.


\subsection{Radio and X-ray properties}\label{sec:data:rx}

\citet{Simpson2006} found this object to be a radio source (VLA J2000
02:17:37.21 -05:13:27.96 ; SXD J2000 02:17:37.21 -05:13:28.0). Its environment
was investigated by \citet{geach021737}. The lens galaxy inhabits a massive
galaxy group (or equivalently a low mass galaxy cluster) at $z_{\rm grp} =
0.648\pm0.001$, with an optical spectroscopic  velocity dispersion
$\sigma^{\rm spec}_{\rm grp} = 770 \pm 170 \kms$. They also quote a group  X-ray
temperature of $k T_X= 1.97\mypm{1.00}{0.59} \kev$, and a group X-ray 
luminosity of $L_{X,\ {\rm [0.3,10\,keV]}} = 1.79\mypm{0.29}{0.83} \times
10^{43}\, {\rm erg/s}$. \citet{geach021737} give the XMM map that allows us to
set the centroid of the group extended X-ray emission to be
$\alpha_{J2000}$=02:17:33.534, $\delta_{J2000}$=-05:13:14.93, with an accuracy
of $\pm10\arcsec$: the group centre is therefore $56\pm10\arcsec$ from \lname.
The $L_X-\sigma$ relation for groups of galaxies (see Fig.~5 in
\citet{geach021737}) shows that the measured group velocity dispersion is
somewhat  higher than one would expect from its X-ray luminosity:  the
$L_X-\sigma$ relation predicts 
$\sigma^{\rm X}_{\rm grp} = 520 \pm 120 \kms$. We can
notice on Fig.~6 of \citet{geach021737} that the positive and negative radial
velocities of the group members do not seem to be distributed randomly across
the group, but are rather distributed in a symmetric way on each side of the
X-ray group centre. This may be a hint that the group is  not relaxed, and
hence  that the measured spectroscopic velocity dispersion might best be taken
as an upper limit on the depth of the potential well. We note that
\citet{geach021737} also identified the lensing nature of the system and
measured both the lens and doubly-imaged  background source redshifts, finding
$\zl=0.646$ and $\zs=1.847$.


\subsection{HST observations}\label{sec:data:hst}

\begin{figure}[htb]
  \centering
  \includegraphics[height=9cm,angle=-90]{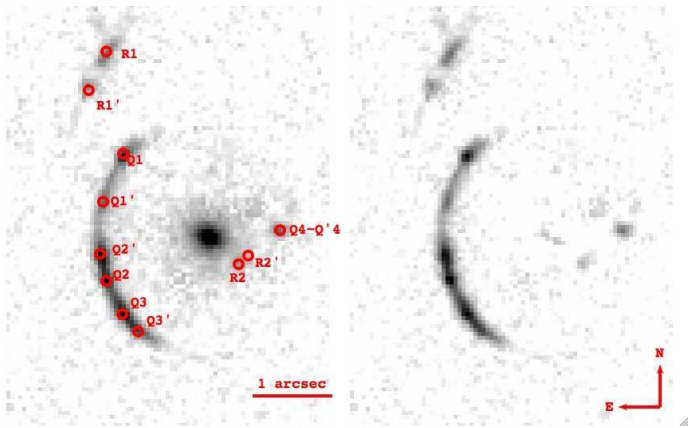}
  \caption{HST image (F606W band) of \lname. 
  The lens redshift is $\zl=0.6459$. 
  The main conjugate points Q1,Q2,Q3,Q4 forming the cusp arc 
  (at redshift $\zs=1.847$) are shown as well as the additional 
  double-image system R1-R2.}\label{021737hst}
\end{figure}

\lname~was observed with the Hubble Space Telescope
Advanced Camera for Surveys (HST/ACS) in the F814W and F606W bands,
as part of our SL2S follow-up snapshot program (SNAP10876, PI Kneib).
A standard reduction was done with {\tt multidrizzle} and we used
{\tt LACosmic} for cosmic ray removal. This method happened
to be more efficient in the F814W band image which is made of $2\times 400$s
exposures, but gave satisfying results in the single $1\times 400$s
F606W exposure. Fig.~\ref{021737hst} shows an image of \lname~in this latter
band, with lensed features labelled (see Sect.~\ref{ssec:multima}).

\subsection{Keck I spectroscopy}\label{ssecKECKdata}

We obtained a spectrum of \lname~with LRIS spectrograph on the Keck I
telescope during the night of December 23rd 2006.
Observations were made in long slit mode ($1\farcs5$ wide) and
oriented East-West.  The spectrum therefore presented two parallel
tracks that allowed us to measure at the same time the lens galaxy
redshift $\zl=0.6459\pm0.0003$ (based on the typical absorption
features like Ca{\sc II} H \& K, G band, Mgb) and the lensed arc redshift
$\zs=1.847$, using the strong emission lines Ly$\alpha$, C{\sc IV},
He{\sc II}, O{\sc III]}, and C{\sc III]}.  Most of them are typical of
a narrow line type II AGN \citep[{\it e.g.}~][]{steidel02}.  This
suggests that the radio source and associated active nucleus,
identified by \citet{Simpson2006} and \citet{geach021737}, might
actually not be at the group redshift $z_{\rm grp}\simeq \zl$ but
rather at redshift $1.847$.  High resolution VLBI imaging would be
able to demonstrate whether the radio emission is consistent with
being lensed by the foreground galaxy.


\section{Lens galaxy properties}\label{sec:LensPpty}

The HST imaging allows us to make precise shape measurements of the
lens galaxy optical surface brightness.  Using the {\tt galfit}
facility \citep{galfit}, we found that the lens galaxy is well fit by
a S\'ersic surface mass density profile (assuming constant stellar
mass-to-light ratio) of the form
\begin{equation}\label{eq:sersic}
\Sigma(R) = \Sigma_e\,\exp\left[-b_n\,\left((R/\Reff)^{1/n}-1\right)\right]\;,
\end{equation}
with S\'ersic index $n\simeq5.2$ (corresponding to $b_n\simeq10.07$)
and effective radius $\Reff$. We thus measured total model magnitudes
$F606W=20.90 \pm 0.05$ and $F814W=19.71 \pm 0.01$ and effective radii
$\Reff=0\farcs90 \pm 0\farcs10 \simeq 6.20 \pm 0.7 \hmkpc$ in the
F814W band.  In addition, {\tt galfit} yielded the axis ratio
$q_*=b/a=0.87\pm0.05$, such that the ellipticity $\epsilon_*=0.14\pm
0.06$ where $\epsilon \equiv (1-q^2)/(1+q^2)$, and the position angle
${\rm PA}_*=78\pm6 ^\circ$ (North to East, ccw)\footnote{Note that
systematic uncertainties dominate the {\tt galfit} run.  We use the
difference between estimates in the F814W and F606W bands as an error
estimate for both $q_*$ and ${\rm PA}_*$}. The measured flux in F814W
band was converted into an absolute rest frame V band luminosity of
$L_V=(1.36\pm0.08) \times 10^{11} \hmlsun$. This includes corrections
from Galactic extinction \citep{schlegel98}, the small filter shifting
correction to translate $z=0.6459$ F814W into a rest frame V band, in
nearly the same way as \citep{gavazzi07} except that, here we do not
apply any passive evolutionary correction to the reference redshift
zero epoch. To do so, one could use the relation constrained by
Fundamental Plane evolution studies \citep{treu01,treu06}:
\begin{equation}\label{eq:dlogmlv}
  \frac{\der \log \frac{M_*}{L_V}  }{\der z} \simeq -0.40 \pm 0.05\;.
\end{equation}

CFHT photometry within a $1\arcsec$ radius aperture around the lens
centre allows us to estimate the corresponding cylindrical stellar
mass. Note that we applied small $\sim 4\%$ corrections to compensate
aperture flux losses due to ground-based seeing (ranging from
$1\farcs1$ in u* to $0\farcs6$ in i').  CFHT magnitudes were also
corrected for Galactic extinction.  Using the publicly available code
\hyperz~\citep{Bolzonella2000} with a Chabrier Initial Mass Function
(IMF) \citep{Chabrier2005} and an intrinsic absorption coefficient
$A_V=0$ typical for nearby early-type galaxies, we obtain a stellar
mass of
$M_*(<1\arcsec) = 1.4 \mypm{0.6}{0.2}\times10^{11}M_{\sun}$.  Using a
Salpeter IMF and/or a non-zero $A_V$ up to $0.3$ (unlikely for such a
massive early-type galaxy), the inferred mass reaches
$M_*(<1\arcsec)\simeq1.7\times10^{11}M_{\sun}$. It results that,
although dominated by stellar evolution systematics, the stellar mass
of the lens galaxy within a
one-arcsec radius is likely in the range
$1.2-2.3\times10^{11}M_{\sun}$.  Using the S\'ersic profile measured
on HST data, we can extrapolate this aperture mass measurement and
estimate a total {\it ``photometric''} stellar mass $M_*=2.1-4.1\times
10^{11}M_{\sun}$.  This value corresponds to a V-band stellar
Mass-to-light ratio
$M_*/L_V=1.5-3.0\,(M/L)_\odot$ at $\zl=0.646$. 
Accounting for luminosity evolution as described
above (Eq.~\ref{eq:dlogmlv}), we get an equivalent zero-redshift 
$M_*/L_V=2.8-5.5\, (M/L)_\odot$,
which is consistent with, although slightly lower than, local estimates
\citep{gerhard01,trujillo04}, but is in good agreement with the strong+weak
gravitational lensing constraints of \citet{gavazzi07}, who found an average
$M_*/L_V=3.84\pm0.40$ in SLACS lens galaxies, once they were 
evolved to redshift zero using Eq.~(\ref{eq:dlogmlv}).


\section{Strong Lensing Analysis}\label{sec:model}

In this section, we present our strong lensing analysis.  It was
performed using the latest version of the \lenstool~code
\citep{jullo07}, which uses a Markov Chain Monte Carlo sampler to
characterise the posterior PDF for the model parameters.


\subsection{Identification of multiple images}\label{ssec:multima}

\lenstool~relies on the identification of conjugate multiple images of
the same background source. For a source imaged $N$~times, since the
unlensed source position is unknown, the location of each of these
points in the image plane results in $2(N-1)$ independent
constraints. When a source is structured in such a way that multiple
sub-components (surface brightness peaks) can be identified, one can
significantly increase the amount of information provided by the
lensing configuration.

For the HST/ACS images we are considering here, we find the typical
astrometric uncertainties on the positions of identified conjugate
image features to be $0\farcs03$. These constraints are summarised in
Table~\ref{tab:multiplets}, while the subsections below detail the way
in which multiple images are identified.

\begin{table}
  \caption{Positions of multiply-imaged, conjugate image features 
  used in the \lenstool~analysis. Units are
  arcseconds West and North relative to the lens galaxy centre. Uncertainties
  are set to a typical value of $0\farcs03$. Column ``\# const'' gives the
  contribution to the total amount of constraints supplied to the
  \lenstool~model.}
  \label{tab:multiplets}
  \centering
  \begin{tabular}{lccccc}
    \hline\hline
      &  Image~1    & Image~2      & Image~3      & Image~4     & \# const\\
  Q   & -1.06, 1.03 & -1.29, -0.53 & -1.06, -0.98 & 0.88, 0.08  &     6\\
  Q'  & -1.33, 0.41 & -1.34, -0.25 & -0.90, -1.18 & 0.88, 0.08  &     6\\
  R   & -1.28, 2.31 &  0.39, -0.33 &              &             &     2\\
  R'  & -1.50, 1.89 &  0.45, -0.25 &              &             &     2\\
total &  & & & & 16 \\ 
    \hline
  \end{tabular}
\end{table}

The HST/ACS image in Fig.~\ref{021737hst} clearly reveals that
\lname~is causing the formation of a tangential arc (Q1-Q2-Q3, in a
typical cusp configuration) along with its small and faint
counter-image (Q4) on the opposite side of the lens, about twice as
close to the lens centre.  The arc is bright
(F606W$_{\rm{arc}}=22.04\pm0.16$ and F814W$_{\rm{arc}}=22.19\pm0.13$).
A rough estimate of its length-to-width ratio~$r$ yields $r\simeq 8$,
which means that the unlensed source magnitudes are $\sim 2.3$ mag
dimmer. The arc has a well-extended surface brightness distribution
resulting from the merging of 3 conjugate multiple images separated by
the critical line. The source thus has to lie across the caustic with
parts of the source pointing outside the tangential cusp. Those parts
of the source will only be imaged 3 times, whereas the others will be
imaged 5 times\footnote{In practice, this odd number of conjugate
images reduces to 2 and 4 multiple images respectively because one of
the images (the central one) is highly demagnified and almost never
observed}.  This configuration makes the identification of conjugate
images of substructures along the arc quite difficult {\it ex nihilo}.

We approach this problem in two stages. In the first stage, we perform
a modelling using the most readily identified (and therefore most
robust) conjugate points in the arc and its counter-image only. This
provides a satisfactory model which we then use for testing further
the more difficult conjugate images along the arc, finally reaching
the astrometric precision of $0\farcs03$ in the predicted image
points.

An interesting feature of \lname~is that a highly-elongated arclet
shape is present at about $2\farcs5$ North-East of the lens galaxy
(labelled R1 in Fig.~\ref{021737hst}). A photometric redshift of this
system $\zrad = 2.90\mypm{0.18}{0.24}$ was obtained from the CFHTLS
photometry
with \hyperz.  The lens modelling of the tangential system Q1-Q2-Q3-Q4
predicts that R1, at that redshift, should be multiply-imaged. Indeed,
we do observe a demagnified counter-image R2 (clearly in the F606W
image, and at slightly lower signal-to-noise in the F814W image).  R2
is unfortunately too faint and buried in the light of the lens galaxy
for an accurate confirmation of its photometric redshift.  However, in
the HST imaging we were able to measure photometry for R2 and R1,
leading to colour indices $(F814W-F606W)_{\rm R1}=0.19\pm0.08$ and
$(F814W-F606W)_{\rm R2}=-0.15\pm0.85$ that are at least consistent
with one another, although this constraint is relatively poor due to
the extreme faintness of R2 ($F606W=27.15\pm0.54$; R1 is much
brighter, $F606W=24.11\pm0.06$).

We can also estimate the ``geometric redshift'' of the R1-R2 source by
including it in the lens modelling.  We identify two secondary
conjugate brightness peaks R'1-R'2, and use these with R1-R2 to
provide 4 new astrometric constraints. We infer the value of $\zrad$
to be very close to the photometric value.

Throughout the rest of this work, we fix the redshift $\zrad$ of the
double-image system to its photometric redshift estimate $\zrad=2.9$,
noting that the uncertainty on this value is small enough for the corresponding
error in the $D_{\rm ls}/D_{\rm s}$ distance ratio to be negligible.
In addition one might expect a third pair of images R3-R'3 of opposite
parity to R2-R'2 with respect to the radial critical line -- if it
exists. These demagnified counter-images are predicted by our lens
model to form well underneath the elliptical galaxy, thus explaining
why we do not observe them. This non-detection prevents us from
accurately fixing the location of the radial critical line, but the
relative distances between R1-R'1 and R2-R'2 still allow us to extend
the range of radial constraints.

\begin{figure}[htb]
  \centering
  \includegraphics[width=8cm,height=8cm]{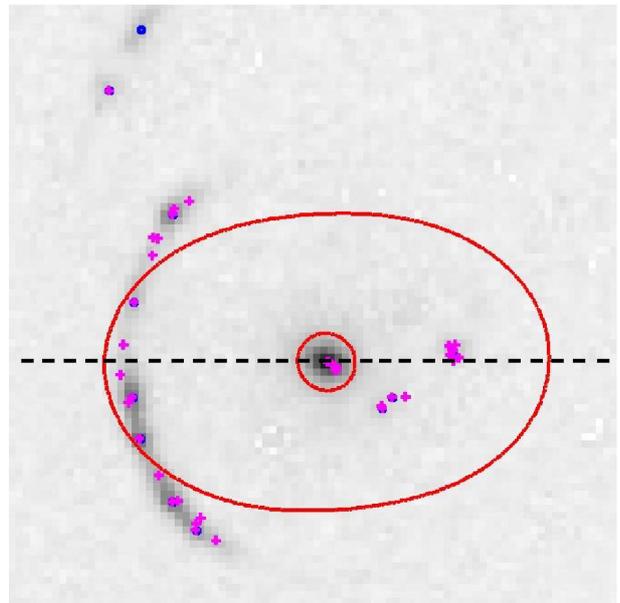}
   \caption{Illustration of the accuracy of the best fit model.
     Small blue circles are input data while red crosses are model-predicted image
     positions after optimisation. Critical curves are shown as red solid
     lines. For reference, the distance of the galaxy centre to the main arc
     is about $1\farcs4$.}.\label{021737draw}
 \end{figure}


\subsection{Model assumptions}\label{sec:method:assumptions}

We now describe our mass model for \lname~and its group environment.
The lens itself is modeled with two components describing the stars in
the lens galaxy, and the dark matter (DM) halo in which they sit. For
the surface density of the former, we assume a S\'ersic profile for
which most model parameters (effective radius $R_e$, index $n$,
ellipticity and orientation) are fixed at the values determined from
the {\tt galfit} model of the HST images. Only the total stellar mass
$M_*$ is taken as a free parameter. Given the total rest-frame V band
luminosity $L_V=(1.36\pm0.08) \times 10^{11} \hmlsun$, one can readily
translate this parameter into a stellar mass-to-light ratio.

The dark matter halo surrounding the lens galaxy is modeled as a
Pseudo Isothermal Mass Distribution
\citep[PIEMD;][]{kassiola93,limousin05}. We also considered a NFW
profile \citep{nfw} as suggested by N-body simulations, and found that
none of our results change significantly either in terms of
goodness-of-fit or regarding the main qualitative trends described in
the results section below.  We fix the centre of the halo to that of
the galaxy light distribution, building on well-established results
from previous galaxy-scale strong lensing studies
\citep[\eg][]{Yoo2006}. (Moreover, when the centroid was freed in our
modelling, we found that the centre of the DM halo has to be $\lesssim
0\farcs1$ from the centre of light.) The orientation $PA_{\rm dm}$,
ellipticity $\epsilon_{\rm dm}$, and velocity dispersion parameter
$\sigma_{\rm dm}$ of the DM halo are let free with broad uniform
priors $0\le \epsilon_{\rm dm}\le 0.6$, $0\le PA_{\rm dm}\le
180^\circ$, and $\sigma_{\rm dm}\le 450 \kms$. We also allow the
possibility of a finite core radius in the range $0<r_{\rm
c,dm}<1\farcs5$. Due to the relatively small projected distance to the
cluster centre ($\simeq 400\kpc$), the galaxy may have a large part of
its DM halo stripped. Lacking observational constraints, we set the
truncation radius $r_{\rm cut}$ to $60\kpc$, an average value
motivated by simulation and galaxy-galaxy lensing results in similarly
high density, cluster environments \citep{priya2,priya3,limousin07a}.

Finally, the host halo of the galaxy group to which \lname~belongs will
have some impact on the lensing properties of the main lens galaxy. To first
order, the perturbing effect of the group  can be characterised by an external
shear, oriented  tangentially 
relative to the group centre. A substantial amount of extra
convergence should also be supplied by this nearby group DM
halo. Since we know where the group is (from the X-ray observations,
Sect.~\ref{sec:data}), we thus decide to model the group by a Singular
Isothermal Sphere (SIS) with velocity dispersion $\sigma_{\rm
grp}$. This model not only predicts the external shear and its {\it
associated} convergence, it also naturally accounts for any higher
order deflections at the lens position.  Typically, if the group
velocity dispersion suggested by the kinematics of group member
galaxies or the intra-cluster gas temperature is correct ($\sigma_{\rm
grp}\sim 630\kms$), one would expect the group to produce external
shear and convergence of order $\vert\gamma_{\rm ext}\vert \sim
\kappa_{\rm ext}\simeq 0.05$ at the location of the lens. We thus
include $\sigma_{\rm grp}$ as the last free parameter of the model,
assigning a uniform prior $0\le \sigma_{\rm grp} \le 900 \kms$ in
order to be able to compare the group mass estimate with the
independent X-ray and spectroscopic values.

Altogether we have a total of 6 free parameters in our lens model.
However, instead of further considering the raw model parameters
above, we introduce the following secondary transformed quantities,
more relevant for the discussion below: the group velocity dispersion
$\sigma_{\rm grp}$, the total stellar mass of the lens galaxy $M_*$,
the dark matter fraction enclosed within the cylinder defined by the
effective radius $f_{\rm dm}(<\Reff)$, the difference in orientations
of the dark matter halo and the stellar component $\Delta PA$ (for
which the errors in the light orientation $PA_*=78\pm 6$ are added in
quadrature to the $PA_{\rm dm}$ estimate) and the ellipticity of the
dark matter halo $\epsilon_{\rm dm}$.


\subsection{Inclusion of more informative priors}\label{sec:model:priors}

We consider two working hypotheses for the modelling of \lname. As our
primary and default strategy, we adopt the most informative prior --
based on the stellar mass derived from the $u^*g'r'i'z'$ aperture
photometry -- i.e. a uniform prior $2.1\le M_*\le 4.1 \times 10^{11}
M_{\sun}$ (hereafter refered to as $M_*$ prior). As a sanity
check, and to investigate the effects of the informative prior, we
also consider the broad uniform priors on the primary parameters
$\sigma_{\rm grp}$, $\epsilon_{\rm dm}$, $PA_{\rm dm}$, $r_{\rm c,dm}$, 
$\sigma_{\rm dm}$ and $M_*$ introduced above. This case will be refered to
as broad prior. Although we will refer primarily to the parameters inferred
from the $M_*$ prior, we will also report for completeness the
parameters inferred from the broad prior. As a further check, we
investigated the assumption that the dark matter halo and the stars
are aligned on the scales probed by the strong lensing features
($6\lesssim R\lesssim12\kpc$).  It turned out that this prior provides
virtually the same constraints as the broad prior case, and
therefore we will not discuss it, for the sake of conciseness.


\section{Results}\label{sec:results}

The lens modelling yields good fits to the positions of all the
multiple images. The recovered model parameters, along with their
$68\%$ confidence level uncertainties and goodness-of-fit --
quantified both in terms of $\chi^2$ per degree of freedom and
Bayesian log-Evidence -- are listed in
Table~\ref{tab:bestfitvalues}. In Fig.~\ref{021737draw}, we show the
excellent matching between the observed positions of the multiple
images (from Table~\ref{tab:multiplets}) and the ones predicted by the
best-fit model.
We stress here that the combination of medians of marginalised
one-dimensional posterior distribution for each parameter (as listed
in Table~\ref{tab:bestfitvalues}) does not necessarily correspond to
the mode of the overall posterior probability distribution that we
nominate {\it ``best fit model''} and approximate with the highest
likelihood MCMC sample.

\subsection{Model parameters}\label{sec:model:results}

\begin{table}[htb]
\caption{Summary of the recovered model parameters. 
Chi-squared values refer to the ``best-fit'' model.
Reported values are the median and the $16\%$ and $84\%$ quantiles 
around it from the marginalised one-dimensional posterior distributions.
Results from the two priors cases are listed for comparison.
Derived constraints on additional secondary parameters are also listed
(slopes are defined in Sect.~\ref{sec:model:profile}).}
\begin{center}
\begin{tabular}{c c c}
\hline\hline
                       &   $M_*$ prior' & broad prior\\
                       &   from photometry & \\
\hline
 constraints             &   16   & 16 \\
 free parameters         &    6   &  6 \\
 $\chi^2$                &   3.1  &  3.5 \\
 $\chi^2$/dof            &   0.31  &  0.35 \\
\hline
   $M_* \; (10^{11}\msun)$     & $3.4\mypm{0.5}{0.7}$ &$6.6\mypm{0.8}{2.1}$ \\
   $\epsilon_{\rm dm}$         & $0.16\mypm{0.20}{0.12}$&$0.27\mypm{0.22}{0.19}$ \\
   $\Delta PA_{\rm dm}$        & $20\mypm{16}{41}$ &$23\mypm{42}{62}$ \\
   $r_{\rm c,dm}$              & $0\farcs48\mypm{0.52}{0.33}$ &$0\farcs72\mypm{0.52}{0.49}$ \\
   $\sigma_{\rm dm}\; (\kms)$  & $218\mypm{43}{28}$ &$106\mypm{85}{72}$ \\
   $\sigma_{\rm grp}\; (\kms)$ & $550\mypm{130}{240}$ &$690\mypm{90}{145}$ \\
\hline
Total mass within $R_{\rm Ein}$  ($10^{11}\msun$) & $5.8\mypm{0.3}{0.4}$ & $7.8\mypm{0.5}{1.1}$ \\
Dark Matter fraction $f_{\rm dm}(<\Reff)$ & $0.41\mypm{0.09}{0.06}$ & $0.06\mypm{0.18}{0.06}$ \\
Group contribution $\vert \gamma_{\rm ext}\vert\sim \kappa_{\rm ext}$ & $0.039\mypm{0.021}{0.026}$ &$0.061\mypm{0.017}{0.023}$   \\
Slope $\alpha$        & $-1.03\mypm{0.22}{0.16}$ & $-1.45\mypm{0.30}{0.16}$ \\
Slope $\alpha^\prime$ & $-1.11\mypm{0.27}{0.18}$ & $-1.68\mypm{0.42}{0.22}$ \\
\hline
\end{tabular}
\label{tab:bestfitvalues}
\end{center}
\end{table}

\begin{figure*}[htb]
\includegraphics[width=\hsize]{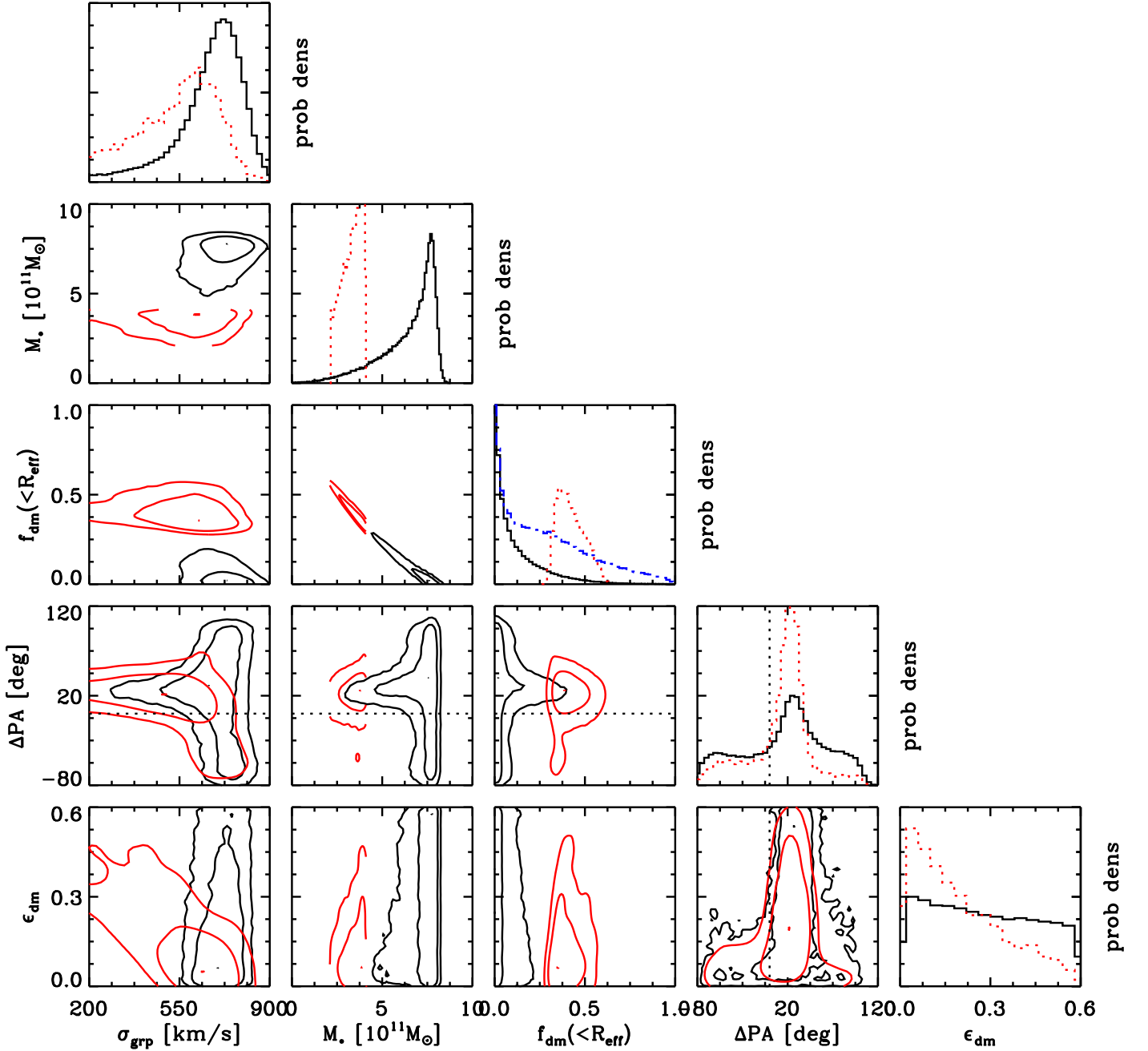}
\caption{Confidence regions for all model parameters, except the lens
  galaxy core radius, which is unconstrained.  For the off-diagonal 2D
  plots, contours represent 1 and 2$-\sigma$ confidence levels.
  Along the diagonal, 1D marginalised distributions are shown.
  Two different prior PDF
  assignments are considered, the basic broad prior (black),
  and the $M_*$ prior (red) based on the \{$u^*g'r'i'z'$\}-band
  CFHTLS photometry and for which stellar mass
  $M_*= 2.1 - 4.1 \times 10^{11}\msun$.
  The dot-dashed blue curve in the $f_{\rm dm}(<\Reff)$ PDF panel represents
  the (un-normalised) implicit broad prior on this derived parameter.
}
\label{fig:dualgroup}
\end{figure*}

We show the marginalised posterior distributions for both the 
$M_*$ and broad prior cases in
Fig.~\ref{fig:dualgroup}. We illustrate pair-wise degeneracies
between the most relevant parameters by plotting the complete (except
for the unconstrained core radius dimension) set of 2D marginalised
posterior PDFs. One-dimensional marginalised posterior distributions
are also shown in the diagonal panels with overlaid, in blue, the {\it
implicit} prior that we assumed for the derived parameters when making
the broad prior assumptions about the primary parameters.
This shows that -- by construction -- our apparently uninformative
broad prior model assumptions will tend to maximise the mass
component that is in the form of stars.

We note the following about our inferences:
\begin{itemize}
\item The misalignment between mass and light $\Delta
  PA=20\mypm{16}{41}$ deg is preferred by the M$_*$ prior. However,
  the same prior also prefers a rounder halo, making the misalignment
  of little physical significance. With the broad prior, the
  misalignment of the dark matter halo and the stellar component
  looses its statistical significance as well.

\item The lens model requires a substantial contribution from the
  group halo: for the $M_*$ prior we infer $\sigma_{\rm
  grp}=550\mypm{130}{240}\kms$  which is consistent with both the
  velocity dispersion of galaxies in the group ($\sigma^{\rm
  spec}_{\rm grp} = 770 \pm 170 \kms$) and the prediction from the
  X-ray luminosity scaling relation ($\sigma^{\rm X}_{\rm grp} = 520
  \pm 120 \kms$).  We observe that the more massive the group, the
  less needed is the contribution of a dark matter halo on the scales
  probed by the Einstein ring. A less massive group halo is more
  in line with $M_*$ prior as we get $\sigma_{\rm grp}=690\mypm{90}{145} \kms$
  for the broad prior. The informative prior thus allows us to obtain tighter
  constraints on the mass of the group.
 
\item The overall group contribution to the mass budget within the
  Einstein radius is about $4\%$, while we find it to contribute about
  $8-13\%$ to the local density at the Einstein radius. The
  implications for the mass budget in the lensing galaxy is thus not
  negligible.  This important issue is discussed in more detail below.

\item The galaxy halo dark matter fraction inside the effective radius
  is highly sensitive to the choice of prior. For the M$_*$ prior we
  obtain $f_{\rm dm}(<\Reff)=0.41\mypm{0.09}{0.06}$. Assuming the
  broad prior would give a much lower value of $f_{\rm
  dm}(<\Reff)=0.06\mypm{0.18}{0.06}$.
  The number obtained with the
  informative $M_*$ prior is consistent with local estimates from the
  SLACS survey \citep[\eg][]{gavazzi07,bolton08b}, where typically $\sim
  30\%$ of the projected mass inside the effective radius is found to be
  in the form of dark matter, independent of any assumption on the
  stellar populations. The broad prior yields a much smaller number,
  which is hard to reconcile with other observations. This lends
  confidence to the use of the informative $M_*$ prior as our primary
  choice.

\item The core radius in the dark matter distribution is unconstrained
  by the data and therefore we do not show the posterior distribution,
  which is indistinguishable from the input prior.
\end{itemize}


\subsection{The mass density profile of  \lname}\label{sec:model:profile}

Thanks to the appearance in \lname~of {\it two sets of
multiple-images} from two sources {\it at different redshifts}, we can
expect this lens, by comparison with the recently discovered double
Einstein ring SDSSJ0946+1006 \citep{gavazzi08}, to provide tight
constraints on the inner parts of its total density profile.
\citet{gavazzi08} showed how important it is to take into account the
multiple deflector plane (``compound lens'') nature of such systems,
since the nearest background source further bends the light rays
coming from the most distant source. In the case of \lname, the two
sources are much further away from the foreground galaxy than in
SDSSJ0946+1006, where the inner ring is at $z=0.609$. Furthermore, the
source positions are not as closely aligned with the lens galaxy as in
SDSSJ0946+1006, where the sources gave rise to almost complete
Einstein Rings. Thus, the impact parameter of the most distant source
in the plane of the intermediate source is much larger than for
SDSSJ0946+1006 in units of the Einstein radius of the intermediate
galaxy (assuming a typical size galaxy). It is therefore not necessary
to treat \lname~as a compound lens. In addition, the image
configuration provides too little information to put constraints on
the mass of the intermediate $\zs=1.847$ galaxy, confirming that its
effect is insignificant for this problem.

\begin{figure}[htb]
 \centering
 \includegraphics[width=9cm,height=7cm]{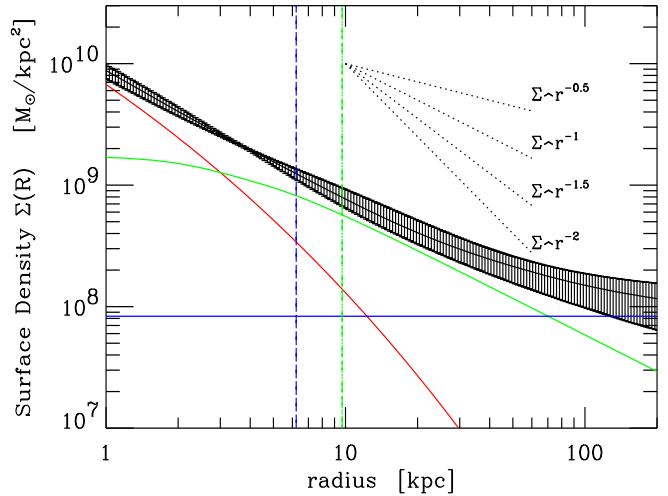}
 \caption{Total surface mass density profile. The effective radius and the 
   Einstein radii are shown as vertical dashed blue
   and dot-dashed green lines (respectively at $\sim 6.2$ and 
   $\sim 10.3\kpc$). The envelope around median values shows the
   $68\%$ confidence interval.
   We show the median stellar density profile (red curve),
   the median DM halo profile (green) and the median effective external
   convergence (blue horizontal line) that successively dominate from
   left to right.}\label{Sigmacompare}
\end{figure}

The nearby galaxy group will provide a substantial amount of
convergence at the lens location, making the issue of the projected
density profile more complicated than for isolated lenses.  In the
previous section we found that about 4\% of the mass enclosed in the
Einstein radius is supplied by the group halo. Locally, at the
Einstein radius, the group halo accounts for $\sim8\%$ of the surface
mass density. Hence the results one gets for the density profile slope
of the lens galaxy strongly depend on the external convergence.  This
issue is well illustrated by Fig.~\ref{Sigmacompare}, which shows the
recovered surface mass density profile: at each radius the density is
evaluated from the median density over all MCMC samples (the envelope
shows the $68\%$ confidence interval, estimated with the $16$ and
$84$~percentiles).  In black is shown the total density profile --
including the stars, the DM halo and the (approximately) constant
convergence from the group.

The red curve shows the contribution of stars, seen to
dominate the density profile on small scales as expected.

\begin{figure}[htb]
 \centering
 \includegraphics[width=\hsize]{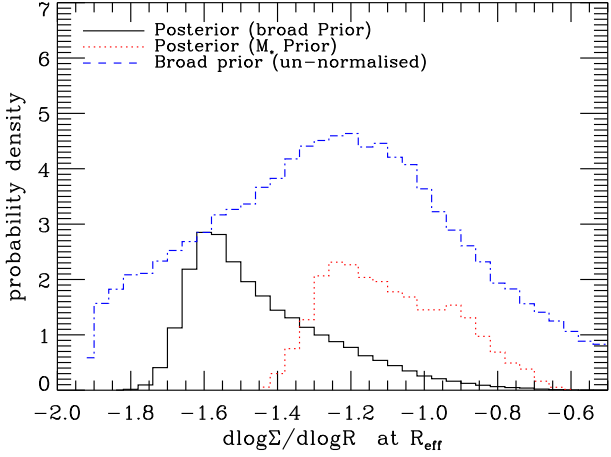}
 \caption{Probability distribution for the logarithmic 
 surface density slope $\alpha$ at the 
 effective radius location for the total (stellar+DM+group convergence)
 density profile. The colour coding for the curves showing 
 posterior probability distribution with different priors is the same as in 
 Fig.~\ref{fig:dualgroup}.}\label{slopescompare}
\end{figure}

Let us now quantify the density profile slope more precisely, focusing
on the logarithmic projected surface mass density slope evaluated at
the effective radius:
\begin{equation}
 \alpha = \left.\frac{\der \log \Sigma_{\rm tot}}
                           {\der \log R}          \right\vert_{R=\Reff}\;.
\end{equation}
This slope parameter would be $-1$ for an isothermal density
profile. We can compute~$\alpha$ for every MCMC sample and thus build
up the posterior PDF $P(\alpha|\data)$.  We plot this distribution in
the top panel of Fig.~\ref{slopescompare}; again, we are showing here
the total profile slope, including dark halo, stellar and group mass
components.

Referring to Fig.~\ref{Sigmacompare}, we can see that out to $\sim4$
effective radii the total density profile (black) is well-approximated
by a single power-law. Fig.~\ref{slopescompare}
shows that this power law index is
$\alpha=-1.03\mypm{0.22}{0.16}$, consistent with the
isothermal case. However, the bottom panel shows the posterior probability 
distribution for the slope parameter $\alpha^\prime$ obtained when we 
{\it remove the external convergence}
from the mass budget. This shifts the result towards a slightly steeper
density profile $\alpha^\prime= -1.11\mypm{0.27}{0.18}$. Constraints on $\alpha$
and $\alpha^\prime$ are listed in table \ref{tab:bestfitvalues} for the broad
and $M_*$ priors.

As in the case of the dark matter fraction $f_{\rm dm}(<\Reff)$, the
constraints we infer on these derived (secondary) parameters need to
be compared to the input priors we assumed. The blue
curve in Fig.~\ref{slopescompare} shows the un-normalised implicit
prior probability distributions for~$\alpha$. This turns out to be
fairly uniform, with only a broad peak close to -1 arising from our
use of the pseudo-isothermal model DM halo (the high Sersic index
stellar mass profile favours slightly steeper-than-isothermal total
density slopes at~$\Reff$).

The difference between the results obtained with the two priors can be
summarised as:
\begin{itemize}
\item the steep density slopes are mostly driven by the high predicted
  stellar mass with the broad prior;
\item a lower stellar mass, as favoured by the informative $M_*$
  prior, implies density slopes that are closer to isothermal with
  $\alpha=-1.03\mypm{0.22}{0.16}$ and $\alpha^\prime=-1.11\mypm{0.27}{0.18}$;
\item the slopes $\alpha$ and $\alpha^\prime$ get closer to one another,
  reflecting the reduction in the contribution to the mass budget from
  the nearby group when the stellar mass is lowered (and the more
  extended DM halo is increased in mass to compensate).
\end{itemize}

\subsection{The external convergence - mass profile degeneracy}\label{sec:model:msd}

Even with two multiply-imaged systems at two different redshifts as we
have here, gravitational lensing alone cannot tell us both the density
profile slope and how much mass is in the form of a uniform external
mass density. However, the degeneracy is broken here, via the
constraints we have placed on the associated external shear provided
by the group halo, the assumption that the group halo has an
isothermal mass distribution, and the functional form assumed for the
mass density profile of the lens galaxy (because the two multiply-imaged
 systems are not exactly at the same projected location in the sky). For
example, by making these assumptions, the external convergence and
shear strength are fixed to be equal.

As we show in Appendix~\ref{sec:appendix}, one could also break this
degeneracy with a measurement of the stellar velocity dispersion of
the lensing galaxy; this would allow us to infer both the density slope and
external convergence at once. We postpone this analysis for future
work. We note that the combination of stellar kinematics of the lens
galaxy in a single aperture and strong lensing constraints coming from
a {\it single multiply-imaged system}
\citep[\eg][]{miralda95,Sand04,treu04,Koopmans06,koopmans06b,czoske08}
cannot break the degeneracy between both density slope and external
convergence by itself: additional information is necessary. This could
include for example spatially resolved stellar kinematics, information
on the external mass distribution, or multiple lens planes together
with assumptions on the form of the mass distribution as in the case
presented here.


\section{Discussion \& Conclusion}\label{sec:modelComp}

In this paper we have  attempted to produce a detailed lens modelling of one
of the most complete systems in the growing ``SL2S'' sample of strong lensing
events in the CFHT Legacy Survey. \lname~is a partial Einstein ring in the
outskirts of an X-ray-emitting group of galaxies. The observational dataset
available for \lname~is quite remarkable, consisting of deep $u^*g'r'i'z'$
CFHTLS photometry, high resolution data from HST/ACS snapshot imaging, a
published XMM X-ray map, the  galaxy velocity dispersion of the host group
\citep[based on 18 radial velocities;][]{geach021737},
and a spectroscopic measurement of the
deflector and source redshifts from Keck/LRIS.

In addition, the system exhibits an extra double-image system, which we have
determined to lie  at higher redshift than the initially detected
arc/counter-arc system. This system gives an important additional handle on
the radial mass distribution. To our knowledge, SDSSJ0946+1006
\citep{gavazzi08} and \lname~constitute the only two known galaxy-scale
systems with multiply-imaged sources at different redshifts.

The influence of the nearby group on the lens configuration is detected as an
external shear that accounts for the slightly misaligned overall potential and
stellar component. 
The lens model including a prior on the stellar mass from the CFHTLS
photometry seems to favour a massive group with velocity
dispersion $\sigma_{\rm grp}=550\mypm{130}{240}\kms$, consistent with 
X-ray luminosity-predicted \citep{geach021737}.
The group environment contribution 
is somewhat degenerate with the mass of the dark matter halo of
the lens galaxy: the higher the group mass, 
the lower the dark matter fraction
in the deflector galaxy. Similarly, if the shear supplied by the group is
reduced, the galaxy dark matter halo and the stellar component must be
preferentially misaligned.

Despite these degeneracies between model parameters, the {\it total}  radial
projected mass profile is well-constrained by the double source plane nature
of \lname. However, we have showed that  the slope of the galaxy
stellar-plus-dark matter  mass density profile at the effective radius differs
according to  whether the convergence supplied by the group is
taken into account in the mass budget or not. 
The effect the group has on the apparent density profile is to make it
shallower by 8\% (from $\alpha^\prime=-1.11$ without the group contribution
to $\alpha=-1.03$ with the group contribution).
Again, this relatively steeper than isothermal profile is consistent with the
lens PG1115+080 in the environment of a group \citep{treu02}.  Statistical
analyses of SLACS lenses also show that they tend to have slightly steeper
density profiles when found in denser environments \citep{auger08} or, more
precisely, when they are satellite of massive halos \citep{treu08}. This
latter result is also supported by simulations of \citet{dobke07} and
\citet{limousin07c}.

As mentioned in the Appendix, a velocity dispersion measurement of the stars
in the central galaxy would help to break the remaining 
degeneracy in this system, tightening the constraints on the galaxy density
profile and the group mass.
We are currently pursuing this observational effort, and we are now working on
the spectroscopic determination of the velocity dispersion of stars in the
lensing galaxy. Likewise, the continuing growth of the SL2S sample will allow
us to tackle the problem of understanding massive galaxy mass distributions
with better statistics.


\begin{acknowledgements}

Based on observations made with the NASA/ESA Hubble Space Telescope,
obtained at the Space Telescope Science Institute, which is operated
by the Association of Universities for Research in Astronomy, Inc.,
under NASA contract NAS 5-26555. These observations are associated
with programs \#10876 and \#11289. Support for programs \#10876 and
\#11289 was provided by NASA through a grant from the Space Telescope
Science Institute.  The authors are thankful to the CFHTLS members and
the Terapix team for their excellent work in reducing and distributing
data to the community. We thank Matthew Auger for insightful
conversations. Part of this project is done under the support of 
National Natural Science
Foundation of China No. 10878003,  10778752, 973Program No. 2007CB815402,
Shanghai Foundation No. 08240514100, 07dz22020, and the Leading
Academic Discipline Project of Shanghai Normal University
(08DZL805). Part
of this work was supported by the Agence Nationale de la Recherche (ANR)
and the Centre National des Etudes Spatiales (CNES). ML acknowledges the
ANR and CNES for its support. The Dark Cosmology centre
is funded by the Danish National Reasearch Fundation. PJM acknowledges
the TABASGO foundation for support in the form of a research
fellowship. TT acknowledges support from the NSF through CAREER award
NSF-0642621, and from the Sloan Foundation through a Sloan Research
Fellowship

\end{acknowledgements}


\bibliographystyle{aa}
\bibliography{references}

\begin{thebibliography}{63}
\expandafter\ifx\csname natexlab\endcsname\relax\def\natexlab#1{#1}\fi

\bibitem[{{Auger}(2008)}]{auger08}
{Auger}, M.~W. 2008, \mnras, 383, L40

\bibitem[{{Auger} {et~al.}(2007){Auger}, {Fassnacht}, {Abrahamse}, {Lubin}, \&
  {Squires}}]{auger07}
{Auger}, M.~W., {Fassnacht}, C.~D., {Abrahamse}, A.~L., {Lubin}, L.~M., \&
  {Squires}, G.~K. 2007, \aj, 134, 668

\bibitem[{{Bolton} {et~al.}(2008{\natexlab{a}}){Bolton}, {Burles}, {Koopmans},
  {Treu}, {Gavazzi}, {Moustakas}, {Wayth}, \& {Schlegel}}]{bolton08a}
{Bolton}, A.~S., {Burles}, S., {Koopmans}, L.~V.~E., {et~al.}
  2008{\natexlab{a}}, \apj, 682, 964

\bibitem[{{Bolton} {et~al.}(2006){Bolton}, {Burles}, {Koopmans}, {Treu}, \&
  {Moustakas}}]{bolton06}
{Bolton}, A.~S., {Burles}, S., {Koopmans}, L.~V.~E., {Treu}, T., \&
  {Moustakas}, L.~A. 2006, \apj, 638, 703

\bibitem[{{Bolton} {et~al.}(2008{\natexlab{b}}){Bolton}, {Treu}, {Koopmans},
  {Gavazzi}, {Moustakas}, {Burles}, {Schlegel}, \& {Wayth}}]{bolton08b}
{Bolton}, A.~S., {Treu}, T., {Koopmans}, L.~V.~E., {et~al.} 2008{\natexlab{b}},
  \apj, 684, 248

\bibitem[{{Bolzonella} {et~al.}(2000){Bolzonella}, {Miralles}, \&
  {Pell{\'o}}}]{Bolzonella2000}
{Bolzonella}, M., {Miralles}, J.-M., \& {Pell{\'o}}, R. 2000, \aap, 363, 476

\bibitem[{{Cabanac} {et~al.}(2007){Cabanac}, {Alard}, {Dantel-Fort}, {Fort},
  {Gavazzi}, {Gomez}, {Kneib}, {Le F{\`e}vre}, {Mellier}, {Pello}, {Soucail},
  {Sygnet}, \& {Valls-Gabaud}}]{cabanac07}
{Cabanac}, R.~A., {Alard}, C., {Dantel-Fort}, M., {et~al.} 2007, \apj, 461, 813

\bibitem[{{Chabrier}(2005)}]{Chabrier2005}
{Chabrier}, G. 2005, in Astrophysics and Space Science Library, Vol. 327, The
  Initial Mass Function 50 Years Later, ed. E.~{Corbelli}, F.~{Palla}, \&
  H.~{Zinnecker}, 41--+

\bibitem[{{Czoske} {et~al.}(2008){Czoske}, {Barnab{\`e}}, {Koopmans}, {Treu},
  \& {Bolton}}]{czoske08}
{Czoske}, O., {Barnab{\`e}}, M., {Koopmans}, L.~V.~E., {Treu}, T., \& {Bolton},
  A.~S. 2008, \mnras, 384, 987

\bibitem[{{Dobke} {et~al.}(2007){Dobke}, {King}, \& {Fellhauer}}]{dobke07}
{Dobke}, B.~M., {King}, L.~J., \& {Fellhauer}, M. 2007, \mnras, 377, 1503

\bibitem[{{Fassnacht} \& {Lubin}(2002)}]{fassnacht02}
{Fassnacht}, C.~D. \& {Lubin}, L.~M. 2002, \aj, 123, 627

\bibitem[{{Fassnacht} {et~al.}(2006){Fassnacht}, {McKean}, {Koopmans}, {Treu},
  {Blandford}, {Auger}, {Jeltema}, {Lubin}, {Margoniner}, \&
  {Wittman}}]{fassnacht06}
{Fassnacht}, C.~D., {McKean}, J.~P., {Koopmans}, L.~V.~E., {et~al.} 2006, \apj,
  651, 667

\bibitem[{{Faure} {et~al.}(2008{\natexlab{a}}){Faure}, {Kneib}, {Covone},
  {Tasca}, {Leauthaud}, {Capak}, {Jahnke}, {Smolcic}, {de la Torre}, {Ellis},
  {Finoguenov}, {Koekemoer}, {Le Fevre}, {Massey}, {Mellier}, {Refregier},
  {Rhodes}, {Scoville}, {Schinnerer}, {Taylor}, {Van Waerbeke}, \&
  {Walcher}}]{faure08a}
{Faure}, C., {Kneib}, J.-P., {Covone}, G., {et~al.} 2008{\natexlab{a}}, \apjs,
  176, 19

\bibitem[{{Faure} {et~al.}(2008{\natexlab{b}}){Faure}, {Kneib}, {Hilbert},
  {Massey}, {Covone}, {Finoguenov}, {Leauthaud}, {Taylor}, {Pires}, \&
  {Scoville}}]{faure08b}
{Faure}, C., {Kneib}, J.-P., {Hilbert}, S., {et~al.} 2008{\natexlab{b}}, ArXiv
  e-prints, astro-ph/0810.4838

\bibitem[{{Gavazzi} {et~al.}(2008){Gavazzi}, {Treu}, {Koopmans}, {Bolton},
  {Moustakas}, {Burles}, \& {Marshall}}]{gavazzi08}
{Gavazzi}, R., {Treu}, T., {Koopmans}, L.~V.~E., {et~al.} 2008, \apj, 677, 1046

\bibitem[{{Gavazzi} {et~al.}(2007){Gavazzi}, {Treu}, {Rhodes}, {Koopmans},
  {Bolton}, {Burles}, {Massey}, \& {Moustakas}}]{gavazzi07}
{Gavazzi}, R., {Treu}, T., {Rhodes}, J.~D., {et~al.} 2007, \apj, 667, 176

\bibitem[{{Geach} {et~al.}(2007){Geach}, {Simpson}, {Rawlings}, {Read}, \&
  {Watson}}]{geach021737}
{Geach}, J.~E., {Simpson}, C., {Rawlings}, S., {Read}, A.~M., \& {Watson}, M.
  2007, \mnras, 873

\bibitem[{{Geiger} \& {Schneider}(1998)}]{geiger98}
{Geiger}, B. \& {Schneider}, P. 1998, \mnras, 295, 497

\bibitem[{{Gerhard} {et~al.}(2001){Gerhard}, {Kronawitter}, {Saglia}, \&
  {Bender}}]{gerhard01}
{Gerhard}, O., {Kronawitter}, A., {Saglia}, R.~P., \& {Bender}, R. 2001, \aj,
  121, 1936

\bibitem[{{Inada} {et~al.}(2008){Inada}, {Oguri}, {Becker}, {Shin}, {Richards},
  {Hennawi}, {White}, {Pindor}, {Strauss}, {Kochanek}, {Johnston}, {Gregg},
  {Kayo}, {Eisenstein}, {Hall}, {Castander}, {Clocchiatti}, {Anderson},
  {Schneider}, {York}, {Lupton}, {Chiu}, {Kawano}, {Scranton}, {Frieman},
  {Keeton}, {Morokuma}, {Rix}, {Turner}, {Burles}, {Brunner}, {Sheldon},
  {Bahcall}, \& {Masataka}}]{inada08}
{Inada}, N., {Oguri}, M., {Becker}, R.~H., {et~al.} 2008, \aj, 135, 496

\bibitem[{{Jullo} {et~al.}(2007){Jullo}, {Kneib}, {Limousin},
  {El{\'{\i}}asd{\'o}ttir}, {Marshall}, \& {Verdugo}}]{jullo07}
{Jullo}, E., {Kneib}, J.-P., {Limousin}, M., {et~al.} 2007, New Journal of
  Physics, 9, 447

\bibitem[{{Kassiola} \& {Kovner}(1993)}]{kassiola93}
{Kassiola}, A. \& {Kovner}, I. 1993, \apj, 417, 450

\bibitem[{{Keeton} {et~al.}(1997){Keeton}, {Kochanek}, \& {Seljak}}]{keeton97}
{Keeton}, C.~R., {Kochanek}, C.~S., \& {Seljak}, U. 1997, \apj, 482, 604

\bibitem[{{King}(2007)}]{king07}
{King}, L.~J. 2007, \mnras, 382, 308

\bibitem[{{Kochanek}(1994)}]{kochanek94}
{Kochanek}, C.~S. 1994, \apj, 436, 56

\bibitem[{{Kochanek}(2002)}]{kochanek02}
{Kochanek}, C.~S. 2002, \apj, 578, 25

\bibitem[{{Kochanek} \& {Blandford}(1991)}]{kochanek91}
{Kochanek}, C.~S. \& {Blandford}, R.~D. 1991, \apj, 375, 492

\bibitem[{{Kochanek} {et~al.}(1999){Kochanek}, {Falco}, {Impey}, {Lehar},
  {McLeod}, \& {Rix}}]{kochanek99}
{Kochanek}, C.~S., {Falco}, E.~E., {Impey}, C.~D., {et~al.} 1999, in American
  Institute of Physics Conference Series, Vol. 470, After the Dark Ages: When
  Galaxies were Young (the Universe at $2<Z<5$), ed. S.~{Holt} \& E.~{Smith},
  163--+

\bibitem[{{Kochanek} \& {Schechter}(2004)}]{kochanek04b}
{Kochanek}, C.~S. \& {Schechter}, P.~L. 2004, in Measuring and Modeling the
  Universe, ed. W.~L. {Freedman}, 117--+

\bibitem[{{Koopmans}(2006)}]{Koopmans06}
{Koopmans}, L.~V.~E. 2006, in Engineering and Science, Vol.~20, EAS
  Publications Series, ed. G.~A. {Mamon}, F.~{Combes}, C.~{Deffayet}, \&
  B.~{Fort}, 161--166

\bibitem[{{Koopmans} {et~al.}(2006){Koopmans}, {Treu}, {Bolton}, {Burles}, \&
  {Moustakas}}]{koopmans06b}
{Koopmans}, L.~V.~E., {Treu}, T., {Bolton}, A.~S., {Burles}, S., \&
  {Moustakas}, L.~A. 2006, \apj, 649, 599

\bibitem[{{Limousin} {et~al.}(2008){Limousin}, {Cabanac}, {Gavazzi}, {Kneib},
  {Motta}, {Richard}, {Thanjavur}, {Foex}, {Crampton}, {Faure}, {Fort},
  {Jullo}, {Marshall}, {Mellier}, {More}, {Pello}, {Soucail}, {Suyu},
  {Swinbank}, {Sygnet}, {Tu}, {Valls-Gabaud}, {Verdugo}, \&
  {Willis}}]{limousin08}
{Limousin}, M., {Cabanac}, R., {Gavazzi}, R., {et~al.} 2008, astro-ph/0812.1033

\bibitem[{{Limousin} {et~al.}(2007{\natexlab{a}}){Limousin}, {Kneib},
  {Bardeau}, {Natarajan}, {Czoske}, {Smail}, {Ebeling}, \&
  {Smith}}]{limousin07a}
{Limousin}, M., {Kneib}, J.~P., {Bardeau}, S., {et~al.} 2007{\natexlab{a}},
  \aap, 461, 881

\bibitem[{{Limousin} {et~al.}(2005){Limousin}, {Kneib}, \&
  {Natarajan}}]{limousin05}
{Limousin}, M., {Kneib}, J.-P., \& {Natarajan}, P. 2005, \mnras, 356, 309

\bibitem[{{Limousin} {et~al.}(2007{\natexlab{b}}){Limousin}, {Sommer-Larsen},
  {Natarajan}, \& {Milvang-Jensen}}]{limousin07c}
{Limousin}, M., {Sommer-Larsen}, J., {Natarajan}, P., \& {Milvang-Jensen}, B.
  2007{\natexlab{b}}, astro-ph/0706.3149

\bibitem[{{Marshall} {et~al.}(2008){Marshall}, {Hogg}, {Moustakas},
  {Fassnacht}, {Bradac}, {Schrabback}, \& {Blandford}}]{marshall08}
{Marshall}, P.~J., {Hogg}, D.~W., {Moustakas}, L.~A., {et~al.} 2008, ArXiv
  e-prints

\bibitem[{{Miralda-Escud\'e}(1995)}]{miralda95}
{Miralda-Escud\'e}, J. 1995, \apj, 438, 514

\bibitem[{{Momcheva} {et~al.}(2006){Momcheva}, {Williams}, {Keeton}, \&
  {Zabludoff}}]{momcheva06}
{Momcheva}, I., {Williams}, K., {Keeton}, C., \& {Zabludoff}, A. 2006, \apj,
  641, 169

\bibitem[{{Myers} {et~al.}(2003){Myers}, {Jackson}, {Browne}, {de Bruyn},
  {Pearson}, {Readhead}, {Wilkinson}, {Biggs}, {Blandford}, {Fassnacht},
  {Koopmans}, {Marlow}, {McKean}, {Norbury}, {Phillips}, {Rusin}, {Shepherd},
  \& {Sykes}}]{myers03b}
{Myers}, S.~T., {Jackson}, N.~J., {Browne}, I.~W.~A., {et~al.} 2003, \mnras,
  341, 1

\bibitem[{{Natarajan} \& {Kneib}(1997)}]{natarajan97}
{Natarajan}, P. \& {Kneib}, J.-P. 1997, \mnras, 287, 833

\bibitem[{{Natarajan} {et~al.}(2002{\natexlab{a}}){Natarajan}, {Kneib}, \&
  {Smail}}]{natarajan02}
{Natarajan}, P., {Kneib}, J.-P., \& {Smail}, I. 2002{\natexlab{a}}, \apjl, 580,
  L11

\bibitem[{{Natarajan} {et~al.}(2002{\natexlab{b}}){Natarajan}, {Kneib}, \&
  {Smail}}]{priya2}
{Natarajan}, P., {Kneib}, J.-P., \& {Smail}, I. 2002{\natexlab{b}}, \apjl, 580,
  L11

\bibitem[{{Natarajan} {et~al.}(2007){Natarajan}, {Kneib}, {Smail}, {Treu},
  {Ellis}, {Moran}, {Limousin}, \& {Czoske}}]{natarajan07}
{Natarajan}, P., {Kneib}, J.-P., {Smail}, I., {et~al.} 2007, ArXiv:0711.4587,
  711

\bibitem[{{Natarajan} {et~al.}(2002{\natexlab{c}}){Natarajan}, {Loeb}, {Kneib},
  \& {Smail}}]{priya3}
{Natarajan}, P., {Loeb}, A., {Kneib}, J.-P., \& {Smail}, I. 2002{\natexlab{c}},
  \apjl, 580, L17

\bibitem[{{Navarro} {et~al.}(1997{\natexlab{a}}){Navarro}, {Frenk}, \&
  {White}}]{navarro1997}
{Navarro}, J.~F., {Frenk}, C.~S., \& {White}, S.~D.~M. 1997{\natexlab{a}},
  \apj, 490, 493

\bibitem[{{Navarro} {et~al.}(1997{\natexlab{b}}){Navarro}, {Frenk}, \&
  {White}}]{nfw}
{Navarro}, J.~F., {Frenk}, C.~S., \& {White}, S.~D.~M. 1997{\natexlab{b}},
  \apj, 490, 493

\bibitem[{{Newton} {et~al.}(2008){Newton}, {Marshall}, \& {Treu}}]{newton08}
{Newton}, E.~R., {Marshall}, P.~J., \& {Treu}, T. 2008, ArXiv e-prints

\bibitem[{{Oguri}(2006)}]{oguri06}
{Oguri}, M. 2006, \mnras, 367, 1241

\bibitem[{{Peng} {et~al.}(2002){Peng}, {Ho}, {Impey}, \& {Rix}}]{galfit}
{Peng}, C.~Y., {Ho}, L.~C., {Impey}, C.~D., \& {Rix}, H.-W. 2002, \aj, 124, 266

\bibitem[{{Sand} {et~al.}(2004){Sand}, {Treu}, {Smith}, \& {Ellis}}]{Sand04}
{Sand}, D.~J., {Treu}, T., {Smith}, G.~P., \& {Ellis}, R.~S. 2004, \apj, 604,
  88

\bibitem[{{Schlegel} {et~al.}(1998){Schlegel}, {Finkbeiner}, \&
  {Davis}}]{schlegel98}
{Schlegel}, D.~J., {Finkbeiner}, D.~P., \& {Davis}, M. 1998, \apj, 500, 525

\bibitem[{{Simpson} {et~al.}(2006){Simpson}, {Mart{\'{\i}}nez-Sansigre},
  {Rawlings}, {Ivison}, {Akiyama}, {Sekiguchi}, {Takata}, {Ueda}, \&
  {Watson}}]{Simpson2006}
{Simpson}, C., {Mart{\'{\i}}nez-Sansigre}, A., {Rawlings}, S., {et~al.} 2006,
  \mnras, 372, 741

\bibitem[{{Steidel} {et~al.}(2002){Steidel}, {Hunt}, {Shapley}, {Adelberger},
  {Pettini}, {Dickinson}, \& {Giavalisco}}]{steidel02}
{Steidel}, C.~C., {Hunt}, M.~P., {Shapley}, A.~E., {et~al.} 2002, \apj, 576,
  653

\bibitem[{{Treu} {et~al.}(2003){Treu}, {Ellis}, {Kneib}, {Dressler}, {Smail},
  {Czoske}, {Oemler}, \& {Natarajan}}]{treu03}
{Treu}, T., {Ellis}, R.~S., {Kneib}, J.-P., {et~al.} 2003, \apj, 591, 53

\bibitem[{{Treu} {et~al.}(2008){Treu}, {Gavazzi}, {Gorecki}, {Marshall},
  {Koopmans}, {Bolton}, {Moustakas}, \& {Burles}}]{treu08}
{Treu}, T., {Gavazzi}, R., {Gorecki}, A., {et~al.} 2008, ApJ accepted,
  astro-ph/0806.1056

\bibitem[{{Treu} {et~al.}(2006){Treu}, {Koopmans}, {Bolton}, {Burles}, \&
  {Moustakas}}]{treu06}
{Treu}, T., {Koopmans}, L.~V., {Bolton}, A.~S., {Burles}, S., \& {Moustakas},
  L.~A. 2006, \apj, 640, 662

\bibitem[{{Treu} \& {Koopmans}(2002)}]{treu02}
{Treu}, T. \& {Koopmans}, L.~V.~E. 2002, \mnras, 337, L6

\bibitem[{{Treu} \& {Koopmans}(2004)}]{treu04}
{Treu}, T. \& {Koopmans}, L.~V.~E. 2004, \apj, 611, 739

\bibitem[{{Treu} {et~al.}(2001){Treu}, {Stiavelli}, {Bertin}, {Casertano}, \&
  {M{\o}ller}}]{treu01}
{Treu}, T., {Stiavelli}, M., {Bertin}, G., {Casertano}, S., \& {M{\o}ller}, P.
  2001, \mnras, 326, 237

\bibitem[{{Trujillo} {et~al.}(2004){Trujillo}, {Burkert}, \&
  {Bell}}]{trujillo04}
{Trujillo}, I., {Burkert}, A., \& {Bell}, E.~F. 2004, \apjl, 600, L39

\bibitem[{{Tu} {et~al.}(2008){Tu}, {Limousin}, {Fort}, {Shu}, {Sygnet},
  {Jullo}, {Kneib}, \& {Richard}}]{tu08}
{Tu}, H., {Limousin}, M., {Fort}, B., {et~al.} 2008, \mnras, 476

\bibitem[{{Williams} {et~al.}(2008){Williams}, {Momcheva}, {Keeton},
  {Zabludoff}, \& {Leh{\'a}r}}]{williams08}
{Williams}, K.~A., {Momcheva}, I., {Keeton}, C.~R., {Zabludoff}, A.~I., \&
  {Leh{\'a}r}, J. 2008, \apj, 672, 733

\bibitem[{{Yoo} {et~al.}(2006){Yoo}, {Kochanek}, {Falco}, \&
  {McLeod}}]{Yoo2006}
{Yoo}, J., {Kochanek}, C.~S., {Falco}, E.~E., \& {McLeod}, B.~A. 2006, \apj,
  642, 22

\end{thebibliography}


\appendix

\section{Constraints on density slope and external convergence with 
three aperture masses}\label{sec:appendix}

In this appendix, we demonstrate that the combination of two multiply-imaged
sources at different redshifts, which define  two projected aperture mass
measurements, and an additional  measurement of a small-scale stellar velocity
dispersion, that can be approximated as a three-dimensional aperture mass
estimate, allow us to constrain simultaneously the density profile slope
parameter {\it and} the external convergence  due to mass lying both along the
line of sight and in the lens plane outside the lens galaxy (but  that does
not affect the lens galaxy's stellar dynamics). 

Let us define a density profile of the form:
\begin{eqnarray}
  \rho(r) &=& \rho_0 (r/r_0)^{-\gamma}\,,\\
  M_{3d}(r) &=& \frac{4 \pi}{3-\gamma} \rho_0 r_0^3  (r/r_0)^{3-\gamma}\,,
\end{eqnarray}
with $1<\gamma<3$. Once projected along the line of sight, we get the
projected surface mass density:
\begin{eqnarray}
  \Sigma(R) &= & \Sigma_0 (R/r_0)^{1-\gamma}= \rho_0 r_0 \sqrt{\pi}\frac{ \Gamma[(\gamma-1)/2]}{\Gamma[\gamma/2]} (R/r_0)^{1-\gamma}\,,\\
  M_{2d}(R) &=&\frac{2 \pi}{3-\gamma} \Sigma_0 r_0^2 (R/r_0)^{3-\gamma}\,.
\end{eqnarray}
We now define a possible constant surface mass density $\Sigma_{\rm sheet}$,
and three mass measurements $M_{2d}^1$, $M_{2d}^2$ and $M_{3d}$ at radii
$R_1$, $R_2$ and $R_3$ that satisfy:
\begin{eqnarray}
  M_{2d}^1 &= &\rho_0 A(\gamma) (R_1/r_0)^{3-\gamma} + \pi R_1^2 \Sigma_{\rm sheet} \\
  M_{2d}^2 &= &\rho_0 A(\gamma) (R_2/r_0)^{3-\gamma} + \pi R_2^2 \Sigma_{\rm sheet} \\
  M_{3d} &= & \rho_0 B(\gamma) (R_3/r_0)^{3-\gamma} \,.
\end{eqnarray}
We can readily see that only in the case of three observables can one
hope to constrain all three of 
$\rho_0$, $\gamma$ and $\Sigma_{\rm sheet}$ ($r_0$
being an irrelevant rescaling here).  
In the equations above we have substituted $A(\gamma)=r_0^3 2 \pi^{3/2}
\Gamma[(\gamma-1)/2]/\Gamma[\gamma/2]/(3-\gamma)$ and $B(\gamma)=r_0^3 4
\pi/(3-\gamma)$\ for brevity. Schematically we thus obtain:
\begin{eqnarray}
\frac{\Gamma[(\gamma-1)/2]}{\Gamma[\gamma/2]} &=&\frac{2}{\sqrt{\pi}}\frac{R_3^{3-\gamma}}{M_{3d}}\frac{ R_2^2M_{2d}^1 - R_1^2M_{2d}^2}{R_2^2R_1^{3-\gamma} - R_1^2 R_2^{3-\gamma}}\,,\label{eq:ap1}\\
\Sigma_{\rm sheet} &=& \frac{1}{\pi}\frac{ M_{2d}^1 R_2^{3-\gamma} - M_{2d}^2 R_1^{3-\gamma}}{ R_1^2 R_2^{3-\gamma} - R_2^2 R_1^{3-\gamma}}\,,\label{eq:ap2}
\end{eqnarray}
where Eq.~(\ref{eq:ap1}) allows us to solve for $\gamma$ and then
Eq.~(\ref{eq:ap2}) yields $\Sigma_{\rm sheet}$.

\end{document}